\documentclass[prb]{revtex4}
\makeatletter

\@addtoreset{equation}{section}

\newcommand{\thesectionnumber}{\@arabic\c@section}

\renewcommand{\theequation}{\thesectionnumber.\@arabic\c@equation}

\makeatother

\usepackage{amsmath,amssymb,graphicx}

\newcommand{\be}{\begin{equation}}
\newcommand{\ee}{\end{equation}}
\newcommand{\Du}{\Delta q_1}
\newcommand{\Dd}{\Delta q_2}

\newcommand{\Dqf}{\Delta q_f}
\newcommand{\Dpu}{\Delta p_1}
\newcommand{\Dpd}{\Delta p_2}
\newcommand{\Dpi}{\Delta p_i}
\newcommand{\Dpf}{\Delta p_f}

\begin{document}
\title{Real trajectories in the semiclassical coherent state propagator}
\author{Marcel Novaes}
\affiliation{Instituto de F\'{i}sica ``Gleb Wataghin",
Universidade Estadual de Campinas, 13083-970 Campinas, S\~ao
Paulo, Brazil}

\begin{abstract}
The semiclassical approximation to the coherent state propagator
requires complex classical trajectories in order to satisfy the
associated boundary conditions, but finding these trajectories in
practice is a difficult task that may compromise the applicability
of the approximation. In this work several approximations to the
coherent state propagator are derived that make use only of real
trajectories, which are easier to handle and have a more direct
physical interpretation. It is verified in a particular example
that these real trajectories approximations may have excellent
accuracy.

\end{abstract}

\maketitle

\section{Introduction}
The path integral representation of the coherent state propagator
$K(z_1,z_2,T)=\langle z_2|e^{-iHT/\hbar}|z_1\rangle$, where the
$|z_i\rangle$ are the usual harmonic oscillator coherent states,
appeared in the works of Klauder and collaborators
\cite{klauder,rubin} and of Weissman.\cite{weiss} A semiclassical,
or stationary phase, approximation leads to classical trajectories
satisfying Hamilton equations of motion but subject to special
boundary conditions that can only be satisfied in a complex phase
space. Aguiar and Baranger also considered this problem
\cite{unpub} and discovered an extra term, which they called
$\mathcal{I}$, in the semiclassical approximation that had been
overlooked in previous studies and that turns out to be essential
for a correct theory \cite{baranger} (the semiclassical spin
propagator has a similar term,\cite{solari} known as the
Solari-Kochetov correction). Numerical calculations involving
complex trajectories in the semiclassical coherent state
propagator have been done for a variety of systems: Adachi
considered a one-dimensional and time-dependent problem with
chaotic dynamics;\cite{adachi} Rubin and Klauder \cite{rubin}, as
well as Xavier and Aguiar,\cite{marcus} have treated $1$D bound
systems; one dimensional tunnelling was considered in
[\onlinecite{xavier}] and also in [\onlinecite{gross}]; Van
Voorhis and Heller presented calculations for one and two
dimensions \cite{voorhis1} and for the $N$-dimensional
Henon-Heiles potential;\cite{voorhis2} Ribeiro {\it et al} have
worked with the $2$D chaotic Nelson potential;\cite{cabelo}
(numerical applications involving the spin coherent states have
also appeared \cite{spin}). Semiclassical approximations based on
complex trajectories for the coordinate wave function, i.e. for
the mixed representation $K(x,z,T)=\langle
x|e^{-iHT/\hbar}|z\rangle$, were also developed, initially for the
one-dimensional case \cite{huber,aguiar,parisio1} and then
generalized to many dimensions.\cite{novaes} The actual
calculation of complex trajectories involves two difficulties:
first, the effective dimensionality of the phase space is doubled,
since both real and imaginary parts of position and momentum must
be computed; second, the boundary conditions are defined part at
initial time and part at final time, and finding the appropriate
classical trajectory becomes a difficult problem known as `root
search'. Therefore approximations that make use only of real
trajectories are certainly desirable.

Since the propagator $K(z_1,z_2,T)$ is a function of time, any
complex trajectory that satisfies the boundary conditions at time
$T$ must belong to a whole `branch' of trajectories, parametrized
by $T$. In general, for a given system and for fixed values of
$z_1$ and $z_2$ there are several such branches. In practice, once
a solution is found for a particular value of $T$, one may obtain
all elements of the same branch by making small steps forward or
backward in time and using appropriate iteractive procedures. It
may happen that for a certain value of time a relevant complex
trajectory has a small (or even null) imaginary part, and in that
case its branch was called `nearly real' by Van Voorhis and
Heller.\cite{voorhis1,voorhis2} It is possible that more than one
`nearly real' branch contribute to the semiclassical propagator
for a given time, and thus one may consider only these branches
and still accurately reproduce interference effects.

A similar analysis can be made for the mixed propagator
$K(x,z,T)$, but in this case one usually holds $T$ fixed and
considers $x$ as a parameter. Varying $x$ thus produces a `family'
of trajectories, and again there may exist several such families.
However, there is always a value of $x$ for which the involved
trajectory is real, and its family was called the `main family' by
Aguiar {\it et al}.\cite{aguiar} Using the main contribution alone
is sometimes a very good approximation, but it can not reproduce
interference because only one trajectory enters the calculation at
a time. On the other hand, as already noted, finding all the
necessary complex trajectories (i.e. performing the `root search')
is usually a difficult problem, specially in more than one
dimension. Therefore the possibility was considered
\cite{aguiar,novaes} of employing only real trajectories in the
semiclassical approximation to $K(x,z,T)$. This was done by
approximating the complex trajectories by real ones, that are
compatible with the quantum uncertainties and satisfy less
restrictive boundary conditions. The final real trajectories
approximations are in principle less accurate than the original
complex one, but they are much simpler and sometimes have
practically the same accuracy.\cite{aguiar,novaes}

The purpose of the present work is to present semiclassical
approximations to $K(z_1,z_2,T)$ that are based only on real
classical trajectories, thus making the calculation much more
tractable. One method that accomplishes exactly this is the
`cellular dynamics', initially developed by Heller \cite{cellular}
(see also [\onlinecite{houches}]) and later generalized and
applied to the stadium billiard with great success.\cite{stadium}
This technique has shown to be accurate even for long
times,\cite{stadium,cel2} and it is actually very close in spirit
to the present work, in the sense that the contribution of a
complex classical trajectory is expanded to second order in the
vicinity of a real one. However, Heller's starting point is the
Van-Vleck-Gutzwiller formula for the semiclassical propagator,
\cite{gutz} while we start from the formulation of Baranger {\it
et al},\cite{baranger} and our results are slightly different from
those of Heller. We also consider a variety of boundary conditions
that the real trajectories may satisfy, something not discussed at
length in.\cite{stadium}

Another approach to the semiclassical coherent state propagator
that is based on real trajectories is the so-called Initial Value
Representations, such as that of Herman and Kluk.\cite{HK} Recent
reviews of this method can be found in [\onlinecite{review}].
Initial value methods are usually easy to apply and reasonably
accurate for long times, but they require a numerical integration
over all possible initial conditions. Since the present method
requires only a few trajectories, at least for short times, it
provides a much clearer physical picture.

This article is divided as follows. In the next section we give a
brief account of the semiclassical approximation to the coherent
state propagator $K(z_1,z_2,T)$ and the complex trajectories. In
section III we present the approximations that are based on the
real trajectories defined by $z_1$ or by $z_2$. Real trajectories
that satisfy mixed boundary conditions are investigated in section
IV. We present an application to a nonlinear oscillator in
sections V and VI and we conclude in section VII.

\section{The semiclassical coherent state propagator}

The coherent states of a harmonic oscillator of mass $m$ and
angular frequency $\omega$ are defined by \be
|z\rangle=\exp\{za^\dag-z^\ast a\}|0\rangle,\ee where $|0\rangle$
is the oscillator ground state. The operators $a^\dag$ and $a$ are
respectively creation and annihilation operators, related to
position $Q$ and momentum $P$ by \be
a=\frac{1}{\sqrt{2}}\left(\frac{Q}{b}+i\frac{P}{c}\right), \quad
a^\dag=\frac{1}{\sqrt{2}}\left(\frac{Q}{b}-i\frac{P}{c}\right).\ee
The parameters $b$ and $c$ define natural scales of the problem,
and are such that $bc=\hbar$ and $c/b=m\omega$. It is easy to see
that if we write \be
z=\frac{1}{\sqrt{2}}\left(\frac{q}{b}+i\frac{p}{c}\right)\ee then
$x$ and $p$ are average values, \be \langle z|Q|z\rangle=q, \quad
\langle z|P|z\rangle=p.\ee The parameters $b$ and $c$ are related
to quantum uncertainties, \be \Delta Q=\frac{b}{\sqrt{2}}, \quad
\Delta P=\frac{c}{\sqrt{2}},\ee and we see that coherent states
are minimum uncertainty states.

These coherent states are never orthogonal, \be \langle
z_2|z_1\rangle=\exp\left\{-\frac{1}{2}|z_1|^2-\frac{1}{2}|z_2|^2+z_1z_2^\ast\right\},\ee
and in the position representation they are Gaussians, \be \langle
x|z\rangle=\pi^{-\frac{1}{4}}b^{-\frac{1}{2}}
\exp\left\{-\frac{(x-q)^2}{2b^2}+\frac{i}{\hbar}p(x-q)\right\}.\ee
In terms of the usual basis of number states $|n\rangle$, defined
such that $a^\dag a|n\rangle=n|n\rangle$, the coherent states may
be written as \be
|z\rangle=e^{-|z|^2/2}\sum_{n=0}^{\infty}\frac{z^n}{\sqrt{n!}}|n\rangle.\ee
It is easy to see that they are eigenstates of the annihilation
operator, $a|z\rangle=z|z\rangle$.

In order to write the semiclassical approximation to the quantum
coherent state propagator \be K(z_1,z_2,T)=\langle
z_2|e^{-iHT/\hbar}|z_1\rangle,\ee we must consider a complex
version of the phase space, i.e. we must make use of a coordinate
$q(t)$ and a momentum $p(t)$ that are complex numbers. Following
the approach of Ref. [\onlinecite{baranger}] we define
\be\label{uv}
u(t)=\frac{1}{\sqrt{2}}\left(\frac{q(t)}{b}+i\frac{p(t)}{c}\right),
\quad
v(t)=\frac{1}{\sqrt{2}}\left(\frac{q(t)}{b}-i\frac{p(t)}{c}\right).\ee
It is of fundamental importance to realize that $v(t)$ is not the
complex conjugate of $u(t)$. In terms of these variables the
boundary conditions become \be\label{boundary} u(0)=u'=z_1, \quad
v(T)=v''=z_2^\ast.\ee There is nothing special about the values
$u(T)=u''$ and $v(0)=v'$, they are to be determined dynamically.
We use hereafter a prime (double prime) to denote initial (final)
values, in order to simplify the formulas and stay close to the
notation of [\onlinecite{baranger}].

The canonical coherent state propagator is \be\label{propag}
K_{\rm sc} (z_1,z_2,T)=\mathcal{N}\sum_{\rm
c.t.}\sqrt{\frac{i}{\hbar}\frac{\partial ^2 S}{\partial u'\partial
v''}}\exp\left\{\frac{i}{\hbar}(S+\mathcal{I})\right\},\ee where
$\mathcal{N}=\exp\{-\frac{1}{2}|z_1|^2-\frac{1}{2}|z_2|^2\}$ is a
normalization factor, the summation is over all classical
trajectories satisfying the boundary conditions, and the complex
action is given by \be S(u',v'',T)=\int_0^T
dt\left[\frac{i\hbar}{2}(\dot{u}v-\dot{v}u)-\mathcal{H}\right]-\frac{i\hbar}{2}(u'v'-u''v'').\ee
This is related to the usual Hamilton action \be
S_H=\int_0^T(p\dot{q}-\mathcal{H})dt\ee by \be\label{sh}
S=S_H-\frac{q'p'-q''p''}{2}-\frac{i\hbar}{2}(u'v'-u''v'').\ee The
Hamiltonian that governs the classical movement according to the
usual Hamilton equations \be
\dot{q}=\frac{\partial\mathcal{H}}{\partial p}, \quad
\dot{p}=-\frac{\partial\mathcal{H}}{\partial q},\ee is the average
value of the quantum Hamiltonian in coherent states, \be
\mathcal{H}=\langle z|H|z\rangle,\ee which is sometimes called the
smoothed Hamiltonian. The quantity $\mathcal{I}$ is related to its
second derivative, \be
\mathcal{I}=\frac{1}{2}\int_0^T\frac{\partial^2
\mathcal{H}}{\partial u
\partial v}dt.\ee

The prefactor in (\ref{propag}) can be written only in terms of
the complex tangent matrix. The classical tangent matrix of a
certain trajectory is the linear application that relates initial
and final displacements about it. We take into account the quantum
uncertainties to define it as follows: \be\label{tangent}
\begin{pmatrix}
  \delta q''/b \\
  \delta p''/c
\end{pmatrix}=\begin{pmatrix}
  m_{qq} & m_{qp} \\
  m_{pq} & m_{pp}
\end{pmatrix}\begin{pmatrix}
  \delta q'/b \\
  \delta p'/c
\end{pmatrix}.\ee
The complex tangent matrix, on the other hand, is defined as \be
\begin{pmatrix}
  \delta u'' \\
  \delta v''
\end{pmatrix}=\begin{pmatrix}
  M_{uu} & M_{uv} \\
  M_{vu} & M_{vv}
\end{pmatrix}\begin{pmatrix}
  \delta u' \\
  \delta v'
\end{pmatrix}.\ee
The relation between the matrix elements of these different
representations is as follows:
\begin{eqnarray} 2M_{uu}=m_{qq}+m_{pp}+im_{pq}-im_{qp},\\
2M_{uv}=m_{qq}-m_{pp}+im_{pq}+im_{qp},\\
2M_{vu}=m_{qq}-m_{pp}-im_{pq}-im_{qp},\\
2M_{vv}=m_{qq}+m_{pp}-im_{pq}+im_{qp}.\end{eqnarray} It is
possible to show that the second derivative of the complex action
is given by \be \frac{i}{\hbar}\frac{\partial ^2 S}{\partial
u'\partial v''}=\frac{1}{M_{vv}},\ee and therefore the
semiclassical coherent state propagator becomes \be
\label{propag2} K_{\rm sc}(z_1,z_2,T)=\sum_{\rm
c.t.}\frac{\mathcal{N}}{\sqrt{M_{vv}}}\exp\left\{\frac{i}{\hbar}(\mathcal{I}+S)\right\}.
\ee

Upon fixing $z_1$ and $z_2$, the squared modulus of this
propagator may be interpreted as a time dependent transition
probability. On the other hand, if we fix $z_1$ and $T$ and
consider $z_2$ as a variable then $|K(z_1,z_2,T)|^2 $ is a phase
space representation, a Husimi function, of the evolved state
$e^{-iHT/\hbar}|z_1\rangle$.

If it happens that $M_{vv}$ tends to zero for a certain
combination of $(z_1,z_2,T)$, then we see that the semiclassical
approximation (\ref{propag2}) diverges. This is called a phase
space caustic \cite{rubin,adachi,voorhis2,aguiar} and the
quadratic approximation used in the derivation of (\ref{propag2})
is not valid in its vicinity. In order to obtain an uniform
approximation that remains valid at caustics it is necessary to
employ a conjugate application of the Bargmann representation, as
discussed in.\cite{uniform} We shall not be concerned with
caustics in this work.

\section{The `leaving' and the `arriving' trajectories}
We have seen that the classical trajectories entering the
semiclassical propagator are determined by mixed boundary
conditions. The initial position and momentum $q'$ and $p'$ are
not the real numbers $q_1$ and $p_1$, but rather some complex
numbers such that $u'=z_1$. Conversely, the final values $q'',p''$
are not $q_2,p_2$ but are such that $v''=z_2^\ast$. It is in
general not a easy task to find such trajectories in practice,
even for simple systems. However, it may happen that the complex
trajectory is close enough to a real one so that we may still
obtain a reasonable result by expanding the propagator to second
order in the vicinity of this real trajectory.\cite{aguiar,novaes}
We investigate this problem with some detail in the next sections.

\subsection{Leaving}
Let us suppose a certain complex classical trajectory that is to
be used in the calculation of the semiclassical propagator, and
let us assume it is not very different from the real trajectory
that starts at the point $(q_1,p_1)$. We call this the `leaving'
trajectory because it leaves the phase space point corresponding
to the initial coherent state. After a time $T$ the position and
the momentum will be some real numbers $(q_f,p_f)$, generally
different from the pair $(q_2,p_2)$. We will expand the complex
action up to second order around this trajectory. If $q'$ is the
initial complex position and $p'$ is the initial complex momentum,
we may write \be q'=q_1+\Du, \quad p'=p_1+\Dpu,\ee where $\Du$ and
$\Dpu$ are assumed to be small (complex) quantities. Moreover, if
$q''$ is the final complex position and $p''$ is the final complex
momentum, we may write in a similar way \be q''=q_f+\Dqf, \quad
p''=p_f+\Dpf.\ee

Therefore we have the approximation
\begin{align}\label{q1p1} S(u',v'',T)&\approx S(z_1,v_r,T)+\left.\frac{\partial
S}{\partial q'}\right|_r\Du+\left.\frac{\partial S}{\partial
p'}\right|_r\Dpu\nonumber\\&+\frac{1}{2}\left.\frac{\partial^2
S}{\partial q'^2}\right|_r\Du^2+\left.\frac{\partial^2 S}{\partial
q'\partial p'}\right|_r\Du\Dpu+\frac{1}{2}\left.\frac{\partial^2
S}{\partial p'^2}\right|_r\Dpu^2,\end{align} where the subscript
$r$ means that the quantity must be evaluated at the real
trajectory (therefore $v'_r=b^{-1}q_1-ic^{-1}p_1$ and
$u''_r=b^{-1}q_f+ic^{-1}p_f$). In order to obtain the derivatives
of the action, we resort to equations (\ref{uv}) and (\ref{sh}).
Noticing that \be \left.\frac{\partial S_H}{\partial
q'}\right|_r=-p_i+\frac{\partial S_H}{\partial q''}\frac{\partial
q''}{\partial q'},\quad \left.\frac{\partial S_H}{\partial
p'}\right|_r=\frac{\partial S_H}{\partial q''}\frac{\partial
q''}{\partial p'},\ee one can obtain the derivatives of the total
action, which are given by \be\label{first}\left.\frac{\partial
S}{\partial q'}\right|_r=
-\frac{ic}{\sqrt{2}}[v'_r+(m_{qq}-im_{pq})u''_r], \quad
\left.\frac{\partial S}{\partial p'}\right|_r=
\frac{b}{\sqrt{2}}[v'_r-(m_{pp}+im_{qp})u''_r].\ee

From the definition of the tangent matrix we have \be
\frac{\partial q''}{\partial q'}=m_{qq}, \quad \frac{\partial
q''}{\partial p'}=\frac{b}{c}m_{qp}, \quad\frac{\partial
p''}{\partial q'}=\frac{c}{b}m_{pq}, \quad\frac{\partial
p''}{\partial p'}=m_{pp},\ee which determines, to first order, the
final differences in terms of the initial ones: \be\label{d2d1}
\Dd=m_{qq}\Du+\frac{b}{c}m_{qp}\Dpu, \quad
\Dpd=\frac{c}{b}m_{pq}\Du+m_{pp}\Dpu.\ee On the other hand, the
boundary conditions \be \frac{q'}{b}+i\frac{p'}{c}=
\frac{q_1}{b}+i\frac{p_1}{c}, \quad \frac{q''}{b}-i\frac{p''}{c}=
\frac{q_2}{b}-i\frac{p_2}{c},\ee provide the secondary relations
\be b^{-1}[\Dd+(q_f-q_2)]=ic^{-1}[\Dpd+(p_f-p_2)], \quad b^{-1}\Du
=-ic^{-1}\Dpu.\ee Solving for $\Du$ and $\Dpu$ in terms of
$(q_f-q_2)$ and $(p_f-p_2)$ we have \be
\Du=-M_{vv}^{-1}[(q_f-q_2)-ib(p_f-p_2)/c], \quad \Dpu=ic\Du/b.\ee
Substituting this in (\ref{q1p1}) one can see that the first order
terms give \be\left.\frac{\partial S}{\partial
q'}\right|_r\Du+\left.\frac{\partial S}{\partial
p'}\right|_r\Dpu=-\frac{i\hbar}{\sqrt{2}}u''_r\left[\frac{q_f-q_2}{b}-i\frac{p_f-p_2}{c}\right]=
-i\hbar u''_r(v''_r-z_2^\ast).\ee

It is easy to take derivatives of equation (\ref{first}) in order
to calculate the quadratic terms. In so doing we neglect
derivatives of the tangent matrix elements, because this would be
a higher order correction. Adding up all quadratic terms and
making the proper identifications, we see that it can be related
to the difference $(v''_r-z_2^\ast)$ as \be \text{quadratic
terms}=-\frac{i\hbar}{2} M_{uv}M_{vv}^{-1}(v''_r-z_2^\ast)^2.\ee
Therefore the final result is the following: \be\label{Kq1p1}
K_{q_1p_1} (z_1,z_2,T)=\frac{\mathcal{N}}{\sqrt{(M_{vv})_r}}
\exp\left\{\frac{i}{\hbar}(\mathcal{I}_r+S_r)+u''_r(v''_r-z_2^\ast)
+\frac{1}{2}\frac{M_{uv}}{M_{vv}}(v''_r-z_2^\ast)^2\right\}.\ee
The subscript in $K_{q_1p_1}$ denotes that this formula was
obtained using the `leaving' trajectory. Notice that the prefactor
and the extra term were not expanded but simply evaluated at the
real trajectory, which is consistent with the original quadratic
derivation of the semiclassical approximation. It is also
important to remember that even though the action $S_r$ is
evaluated at a real trajectory, it continues to be a complex
number.

The expression (\ref{Kq1p1}) depends quadratically on the
difference between the final value of the variable $v$ along the
real trajectory and the value that it would have in the complex
trajectory. If by some reason the situation is such that $v''_r$
and $z_2^\ast$ coincide, then this formula and the original one
(\ref{propag}) will give the same result. One may argue that it is
possible to obtain the same expression by expanding the action as
\be S\approx S_r+\left.\frac{\partial S}{\partial
v''}\right|_r(v''_r-z_2^\ast)+\frac{1}{2}\left.\frac{\partial^2
S}{\partial v''^2}\right|_r (v''_r-z_2^\ast)^2.\ee This is
certainly true and actually an easy calculation. We have chosen
the long way of using the position/momentum variables because this
will be the only possibility in the next section.

\subsection{Arriving}
What we call the `arriving' trajectory is the real trajectory that
starts in a certain initial point $(q_i,p_i)$ and after a time $T$
arrives at the point $(q_2,p_2)$. We can use this trajectory to
approximate the semiclassical propagator in the very same way that
we did with the `leaving' trajectory. Similar to the previous
arguments, we write \be q'=q_i+\Du, \quad p'=p_i+\Dpu, \quad
q''=q_2+\Dd, \quad p''=p_2+\Dpd.\ee Inverting equation
(\ref{tangent}) we see that \be \frac{\partial q'}{\partial
q''}=m_{pp}, \quad \frac{\partial q'}{\partial
p''}=-\frac{b}{c}m_{qp}, \quad\frac{\partial p'}{\partial
q''}=-\frac{c}{b}m_{pq}, \quad\frac{\partial p'}{\partial
p''}=m_{qq}.\ee Using these relations we can write the initial
differences in terms of the final ones, analogously to what we did
in (\ref{d2d1}). Using the boundary conditions it is possible to
show that \be \Dd=-M_{vv}^{-1}[(q_i-q_1)+ib(p_i-p_1)/c], \quad
\Dpd=-ic\Dd/b.\ee The first derivatives of the action are in this
case given by \be \left.\frac{\partial S}{\partial q''}\right|_r=
-\frac{ic}{\sqrt{2}}[u''_r+(m_{pp}-im_{pq})v'_r], \quad
\left.\frac{\partial S}{\partial p''}\right|_r=
-\frac{b}{\sqrt{2}}[u''_r-(m_{qq}+im_{qp})v'_r].\ee

We now expand the complex action to second order around this real
trajectory. After simplifications, we obtain \be \label{Kq2p2}
K_{q_2p_2}
(z_1,z_2,T)=\frac{\mathcal{N}}{\sqrt{(M_{vv})_r}}\exp\left\{\frac{i}{\hbar}(\mathcal{I}_r+S_r)+v'_r(u'_r-z_1)
+\frac{1}{2}\frac{M_{vu}}{M_{vv}}(u'_r-z_1)^2\right\},\ee where
the meaning of the subscript is evident. This time the expression
depends on the difference between the initial value of the
variable $u$ in the real trajectory and the value that it would
have in the complex one. Its interpretation is quite close to that
of (\ref{Kq1p1}).

\section{Other possible real trajectories}
In the previous section we saw that we may expand the
semiclassical propagator in the vicinity of the real trajectories
determined by the initial or by the final labels, $(q_1,p_1)$ and
$(q_2,p_2)$, which we called the `leaving' and the `arriving'
trajectories respectively. Although these are probably the most
natural real trajectories approximations, we can devise four more
possibilities that are also interesting. Of course one may use any
real trajectory to build an approximation --in fact, in principle
it should be possible to find the `best' choice by a variational
approach, but this seems to be a highly nontrivial problem--, but
the idea here is to obtain explicit formulas for the most natural
cases. These are the four trajectories that are determined by
pairwise combination of the coherent state labels.

We shall present a detailed calculation for the case when the
trajectories determined by the pair $(q_1,q_2)$ are used. All
other cases can be treated in a very similar way, and for them we
shall be less explicit.

\subsection{From $q_1$ to $q_2$} Let us consider a trajectory which
satisfies the following boundary conditions: it leaves $q_1$ at
time zero and arrives at $q_2$ at time $T$. Its initial and final
momenta, $p_i$ and $p_f$, remain unknown, but are real numbers.
Differently from the previous section, now there may be more than
one trajectory satisfying these requirements. We write \be
q'=q_1+\Du, \quad q''=q_2+\Dd, \quad p'=p_i+\Dpi, \quad
p''=p_f+\Dpf.\ee The initial and final momenta are regarded as
functions of the initial and final positions. Therefore we may
write \be\label{dpdq} \Dpi=\left.\frac{\partial p'}{\partial
q'}\right|_r\Du+\left.\frac{\partial p'}{\partial
q''}\right|_r\Dd, \quad \Dpf=\left.\frac{\partial p''}{\partial
q'}\right|_r\Du+\left.\frac{\partial p''}{\partial
q''}\right|_r\Dd,\ee where again the subscript $r$ means that the
quantity must be evaluated at the real trajectory. On the other
hand the boundary conditions $u'=z_1$ and $v''=z_2^\ast$ imply
that \be\label{bound} \Dpi=\frac{ic}{b}\Du-(p_i-p_1), \quad
\Dpf=-\frac{ic}{b}\Dd-(p_f-p_2).\ee

Since we are considering $q'$ and $q''$ as independent variables,
the partial derivatives in (\ref{dpdq}) are given by \be
\frac{\partial p'}{\partial q'}=-\frac{c}{b}\frac{m_{qq}}{m_{qp}},
\quad \frac{\partial p'}{\partial
q''}=\frac{c}{b}\frac{1}{m_{qp}}, \quad \frac{\partial
p''}{\partial q'}=-\frac{c}{b}\frac{1}{m_{qp}}, \quad
\frac{\partial p''}{\partial
q''}=\frac{c}{b}\frac{m_{pp}}{m_{qp}},\ee where we have used that
$m_{qq}m_{pp}-m_{qp}m_{pq}=1$. Substituting this in (\ref{dpdq})
and using (\ref{bound}) we have \be\label{dqq}
\frac{\Du}{b}=\frac{m_{qp}}{c}\frac{[(p_f-p_2)-M_2(p_i-p_1)]}{1-M_1M_2},
\quad
\frac{\Dd}{b}=\frac{m_{qp}}{c}\frac{[M_1(p_f-p_2)-(p_i-p_1)]}{1-M_1M_2},\ee
where we have defined the complex numbers \be M_1=m_{qq}+im_{qp},
\quad M_2=m_{pp}+im_{qp}.\ee

We now expand the complex action around this real trajectory up to
second order,
\begin{align}\label{sqq} S&\approx
S_r+\left.\frac{\partial S}{\partial
q'}\right|_r\Du+\left.\frac{\partial S}{\partial
q''}\right|_r\Dd\nonumber\\&+\frac{1}{2}\left.\frac{\partial^2
S}{\partial q'^2}\right|_r\Du^2+\left.\frac{\partial^2 S}{\partial
q'\partial q''}\right|_r\Du\Dd+\frac{1}{2}\left.\frac{\partial^2
S}{\partial q''^2}\right|_r\Dd^2.\end{align} Noticing that \be
\left.\frac{\partial S_H}{\partial q'}\right|_r=-p_i,\quad
\left.\frac{\partial S_H}{\partial q''}\right|_r=p_f,\ee we can
obtain the derivatives of the total action, \be
\left.\frac{\partial S}{\partial
q'}\right|_r=\frac{c}{\sqrt{2}m_{qp}}[u''_r-v'_r(m_{qq}+im_{qp})],
\quad  \left.\frac{\partial S}{\partial
q''}\right|_r=\frac{c}{\sqrt{2}m_{qp}}[v'_r-u''_r(m_{pp}+im_{qp})].\ee
After simplifications, the linear terms can be written as \be
\text{linear
terms}=-\frac{b}{\sqrt{2}}[v'(p_i-p_1)-u''(p_f-p_2)].\ee

We now calculate the second derivatives and substitute (\ref{dqq})
in (\ref{sqq}). After many simplifications, the final result can
be shown to be
\begin{align}\label{kqq} K_{q_1q_2}(z_1,z_2,T)&=\sum_{\rm c.t}\frac{\mathcal{N}}{\sqrt{(M_{vv})_r}}
\exp\left\{\frac{i}{\hbar}(\mathcal{I}_r+S_r)+\frac{iz_2}{\sqrt{2}c}(p_f-p_2)
-\frac{iz_1^\ast}{\sqrt{2}c}(p_i-p_1)\right\}\nonumber\\
&\times\exp\left\{-\frac{A_1}{2c^2}(p_i-p_1)^2-\frac{A_2}{2c^2}(p_f-p_2)^2
-\frac{A_{12}}{2c^2}(p_i-p_1)(p_f-p_2)\right\},\end{align} where
\be A_1=1-\frac{1}{2}\left(\frac{1-M_1^\ast M_2}{1-M_1M_2}\right),
\quad A_2=1-\frac{1}{2}\left(\frac{1-M_2^\ast
M_1}{1-M_1M_2}\right), \quad A_{12}=\frac{2im_{qp}}{1-M_1M_2}.\ee

This expression is more complicated that the ones we obtained in
section III. This is so because the classical trajectories
involved are determined by mixed boundary conditions, i.e. their
initial and final positions. Its structure is nevertheless still
the same: it depends on differences between the values of the
variables in the real trajectories and the corresponding coherent
state labels. The most important property of this formula is that
the initial momentum $p_i$ is not known {\it a priori}. It must be
determined as a function of the given parameters, and in fact
there may be more than one possible value for it. Notice that
since $p_i$ and also $p_f$ depend nontrivially on $z_1$, $z_2$ and
$T$ this formula is not a simple Gaussian as it may seem at first.
Once again, even though the function $S_r$ is evaluated at a real
classical trajectory, it will in general be a complex number.

Notice that the differences $p_i-p_1$ and $p_f-p_2$ are always
divided by the momentum uncertainty $c$. Therefore only classical
trajectories whose initial momentum is within a distance $c$ from
$p_1$ may be important for the semiclassical propagator. The same
reasoning applies to the final momentum. We see that the real
trajectories to be used in this formalism must be compatible with
the quantum uncertainty principle.

As a simple illustration of this formula, let us consider a
harmonic oscillator of unit mass and angular frequency
$\omega=c/b$. An initial condition $(q',p')$ leads, after a time
$T$, to the final values \be q''=q'\cos(\omega
T)+\frac{p'}{\omega}\sin(\omega T), \quad p''=-\omega
q'\sin(\omega T)+p'\cos(\omega T).\ee If we impose that the
trajectory must start in $q_1$ and end in $q_2$ then it is easy to
see that there is only one possibility that satisfies these
boundary conditions, for which \be
p_i=\frac{\omega(q_2-q_1\cos(\omega T))}{\sin(\omega T)}, \quad
p_f=\frac{\omega(q_2\cos(\omega T)-q_1)}{\sin(\omega T)}.\ee In
this case we have $m_{qp}=\sin(\omega T)$ and $M_1=M_2=e^{i\omega
T}$, which leads to $A_1=A_2=1$ and $A_{12}=-e^{-i\omega T}$. The
$e^{i\mathcal{I}_r/\hbar}$ term cancels the prefactor. Finally,
using that $S_r=e^{-i\omega T}(q_1/b+ip_i/c)(q_2/b-ip_f/c)/2i$ we
obtain \be
K_{q_1q_2}(z_1,z_2,T)=\exp\{-\frac{1}{2}(|z_1|^2+|z_2|^2)+e^{-i\omega
T}z_1z_2^\ast\},\ee which is precisely the exact result. This
comes as no surprise since the exact action in this case is of
second order to begin with and thus all semiclassical
approximations we consider in this work will be exact.

\subsection{From $q_1$ to $p_2$}
We now consider the real trajectory that starts in $q'=q_1$ with a
certain momentum $p'=p_i$ and, after a time $T$, is in a final
point $q''=q_f$ with the momentum $p''=p_2$. We therefore treat
$q'$ and $p''$ as independent variables, in which case we have the
following partial derivatives: \be \frac{\partial p'}{\partial
q'}=-\frac{c}{b}\frac{m_{pq}}{m_{pp}}, \quad \frac{\partial
p'}{\partial p''}=\frac{\partial q''}{\partial
q'}=\frac{1}{m_{pp}}, \quad \frac{\partial q''}{\partial
p''}=\frac{b}{c}\frac{m_{qp}}{m_{pp}}.\ee We may calculate the
action's first derivatives, \be \left.\frac{\partial S}{\partial
q'}\right|_r=-\frac{ic}{\sqrt{2}m_{pp}}[u''_r+v'_r(m_{pp}-im_{pq})],
\quad  \left.\frac{\partial S}{\partial
p''}\right|_r=\frac{b}{\sqrt{2}m_{pp}}[v'_r-u''_r(m_{pp}+im_{qp})],\ee
and after writing \be q'=q_1+\Du, \quad q''=q_f+\Dqf, \quad
p'=p_i+\Dpi, \quad p''=p_2+\Dpd,\ee we may also obtain, using an
expansion analogous to (\ref{dpdq}) and the boundary conditions,
the relations
\begin{eqnarray}
\frac{\Du}{b}=-\frac{im_{pp}}{c}\frac{[M_2(p_i-p_1)-ic(q_f-q_2)/b]}{1+M_2M_3^\ast},
\\
\frac{\Dpd}{c}=-\frac{im_{pp}}{b}\frac{[M_3(q_f-q_2)-ib/c(p_i-p_1)]}{1+M_2M_3^\ast},\end{eqnarray}
where $M_2$ has already been defined and $M_3=m_{pp}+im_{pq}$.

After calculating the action's second derivatives, the final
result is
\begin{align}\label{kq1p2} K_{q_1p_2}(z_1,z_2,T)&=\sum_{\rm
c.t}\frac{\mathcal{N}}{\sqrt{(M_{vv})_r}}
\exp\left\{\frac{i}{\hbar}(\mathcal{I}_r+S_r)-\frac{z_2}{\sqrt{2}b}(q_f-q_2)-
\frac{iz_1^\ast}{\sqrt{2}c}(p_i-p_1)\right\}\nonumber\\
&\times\exp\left\{
-\frac{B_1}{2c^2}(p_i-p_1)^2-\frac{B_2}{2b^2}(q_f-q_2)^2+\frac{B_{12}}{\hbar}(p_i-p_1)(q_f-q_2)\right\},\end{align}
where the coefficients are given by \be
B_1=1-\frac{1}{2}\left(\frac{1-M_2M_3}{1+M_2M_3^\ast}\right),
\quad B_2=1-\frac{1}{2}\left(\frac{1-M_2^\ast
M_3^\ast}{1+M_2M_3^\ast}\right), \quad
B_{12}=\frac{im_{pp}}{1+M_2M_3^\ast}.\ee We see that the
semiclassical propagator obtained is quite similar in structure to
the one presented in the previous subsection. Only this time we
have position and momentum in a more equal footing. As $p_i-p_1$
is always divided by $c$ and $q_f-q_2$ is always divided by $b$,
we see that again the quantum uncertainties play a fundamental
role in selecting the relevant classical trajectories.

\subsection{From $p_1$ to $q_2$}
It is also possible to fix the initial momentum as $p_1$ and then
search for an initial position $q_i$ such that the final position
is $q_2$. In that case the final momentum will be some $p_f$.
Proceeding in complete analogy with the previous cases, we take
$p'$ and $q''$ to be independent variables and calculate
derivatives of $q'$, $p''$ and $S$ with respect to them. After
obtaining the values of $\Dpu$ and $\Dd$ in terms of $(q_i-q_1)$
and $(p_f-p_2)$ and expanding the action to second order, the
final result will be \begin{align}
K_{p_1q_2}(z_1,z_2,T)&=\sum_{\rm
c.t}\frac{\mathcal{N}}{\sqrt{(M_{vv})_r}}
\exp\left\{\frac{i}{\hbar}(\mathcal{I}_r+S_r)+\frac{iz_2}{\sqrt{2}c}(p_f-p_2)-
\frac{z_1^\ast}{\sqrt{2}b}(q_i-q_1)\right\}\nonumber\\
&\times\exp\left\{
-\frac{C_1}{2b^2}(q_i-q_1)^2-\frac{C_2}{2c^2}(p_f-p_2)^2-\frac{C_{12}}{\hbar}(q_i-q_1)(p_f-p_2)\right\},\end{align}
where the coefficients are given by \be
C_1=1-\frac{1}{2}\left(\frac{1-M_1^\ast
M_4^\ast}{1+M_1M_4^\ast}\right), \quad
C_2=1-\frac{1}{2}\left(\frac{1-M_1M_4}{1+M_1M_4^\ast}\right),
\quad C_{12}=\frac{im_{qq}}{1+M_1M_4^\ast},\ee with
$M_4=m_{qq}+im_{pq}$.

\subsection{From $p_1$ to $p_2$}
Finally, we consider the trajectory determined by the pair
$(p_1,p_2)$. This has initial and final positions $q_i$ and $q_f$,
respectively. The procedure to obtain the semiclassical
approximation is certainly clear by now, so it will not be
repeated in any detail. The final result in this case will be
\begin{align} K_{p_1p_2}(z_1,z_2,T)&=\sum_{\rm
c.t}\frac{\mathcal{N}}{\sqrt{(M_{vv})_r}}
\exp\left\{\frac{i}{\hbar}(\mathcal{I}_r+S_r)-\frac{z_1^\ast}{\sqrt{2}b}(q_i-q_1)-
\frac{z_2}{\sqrt{2}b}(q_f-q_2)\right\}\nonumber\\
&\times\exp\left\{
-\frac{D_1}{2b^2}(q_i-q_1)^2-\frac{D_2}{2b^2}(q_f-q_2)^2-\frac{D_{12}}{b^2}(q_i-q_1)(q_f-q_2)\right\},\end{align}
where the coefficients are given by \be
D_1=1-\frac{1}{2}\left(\frac{1-M_3M_4^\ast}{1-M_3^\ast
M_4^\ast}\right), \quad D_2=1-\frac{1}{2}\left(\frac{1-M_3^\ast
M_4}{1-M_3^\ast M_4^\ast}\right), \quad
D_{12}=\frac{im_{pq}}{1-M_3^\ast M_4^\ast}.\ee

\subsection{Summary of section IV}
In this section we have obtained four different semiclassical
approximations to the quantum coherent state propagator that are
based only on real trajectories. The trajectories considered share
the property that they are not determined by initial or final
values, but satisfy mixed boundary conditions. Therefore finding
them in practice is not trivial, but is certainly easier than
finding the original complex ones. All the semiclassical
propagators obtained are in principle able to reproduce quantum
effects such as interference, since there may be more than one
classical trajectory involved. They will be affected by caustics
just like the original formula (\ref{propag2}), but the location
of such caustics will change because $(M_{vv})_r$ is different for
each one of them.

Which one of the several formulas obtained here and in section III
is more accurate will depend on the particular problem at hand. We
have considered only initial and final coherent states with the
same value of the parameter $b$, but a generalization of the
semiclassical propagator was presented \cite{parisio} for more
general $b$'s, and the calculations presented here may be adapted
to that case with no essential difficulty. Let us suppose for a
moment that the initial coherent state $|z_1\rangle$ has a
position uncertainty $b_1$ while $|z_2\rangle$ has a position
uncertainty $b_2$. If these numbers are small that means the
states are very narrow in the position representation, while
having a large uncertainty in momentum. In that case we conjecture
that an approximation in the spirit of $K_{q_1q_2}$ would be the
most effective one. If $b_1$ is small but $b_2$ is large, than
$K_{q_1p_2}$ would be a better candidate, and so on. Of course for
the free particle and the harmonic oscillator they are all exact,
regardless of the values of $b_1$ and $b_2$.

In the next section, we present an application of the formalism
just presented to a nonlinear system. The purpose is not to
attempt an exhaustive investigation of the several possibilities,
but rather to illustrate the method in a simple case. We have
chosen a system for which many analytical results are possible so
that the main properties of the theory do not disappear under
numerical calculations.

\section{Application to a nonlinear oscillator: short time}
We consider the nonlinear Hamiltonian \be\label{kerr}
H=\hbar\omega(a^\dag a)^2
=\frac{1}{\hbar\omega}\frac{(p^2+\omega^2q^2-\hbar
\omega)^2}{4},\ee which is diagonal in the usual number basis, \be
H|n\rangle=E_n|n\rangle=\hbar\omega n^2|n\rangle.\ee The quantum
propagator in this case is quite simple: \be K(z_1,z_2,T)=\langle
z_2|e^{-iHT/\hbar}|z_1\rangle=\mathcal{N}\sum_{n=0}^{\infty}\frac{(z_1z_2^\ast)^n}{n!}e^{-in^2\omega
T}.\ee We shall be interested, for simplicity, only in the
diagonal case \be\label{diag}
K(z_1,z_1,T)=e^{-|z_1|^2}\sum_{n=0}^{\infty}\frac{|z_1|^{2n}}{n!}e^{-in^2\omega
T},\ee whose squared modulus is the return probability, \be
P(z_1,T)=|K(z_1,z_1,T)|^2.\ee This function is periodic with
period $T_r=2\pi/\omega$. In the semiclassical limit the term that
is responsible for the largest contribution to the sum in
(\ref{diag}) is $n_0\approx|z_1|^2$. If we linearize the exponent
in the vicinity of this term we have \be\label{linear}
P(z_1,T)\approx e^{-2|z_1|^2}\left|\sum_{n\approx
n_0}\frac{|z_1|^{2n}}{n!}e^{2in_0n\omega T}\right|^2. \ee Notice
that expression (\ref{linear}) has a distinct time scale, \be
T_c=\frac{\pi}{n_0\omega}.\ee The quantities $T_r$ and $T_c$ are
usually called revival time and classical time.\cite{revivals}

For short times, we can approximate $|K(z_1,T)|^2\approx
1-\left(\langle H^2\rangle-\langle H\rangle^2\right) T^2/\hbar^2$,
where $\langle \cdot\rangle$ denotes an average value in the state
$|z_1\rangle$. For the system in question, this gives
\be\label{short0} P(z_1,T)\approx
1-\left(4|z_1|^6+6|z_1|^4+|z_1|^2\right)\omega^2T^2.\ee Let us
write $z_1=(q+ip)/\sqrt{2}$ and take for simplicity the value
$q=0$. Since the movement in phase space has circular symmetry,
this choice is of no fundamental importance. The short-time
expansion (\ref{short0}) becomes simply \be\label{short}
P(z_1,T)\approx 1-\frac{1}{2}\left(p^6+3p^4+p^2\right)T^2.\ee

Let us now turn to the semiclassical approximation. From now on we
set $\hbar=\omega=1$, which implies $b=c=1$. The smoothed
Hamiltonian associated with (\ref{kerr}) is \be
\mathcal{H}=\frac{(p^2+q^2)(p^2+q^2+2)}{4}=uv(uv+1),\ee and the
corresponding Hamilton equations are \be
\dot{q}=\frac{\partial\mathcal{H}}{\partial p}=\sigma p, \quad
\dot{p}=-\frac{\partial\mathcal{H}}{\partial q}=-\sigma q,\ee
where we have defined $\sigma=p^2+q^2+1$. If we note that
$\{\sigma,\mathcal{H}\}=0$, and thus that $\sigma$ is a constant
of the motion, then it is clear that \be q''=q'\cos(\sigma
t)+p'\sin(\sigma t), \quad p''=p'\cos(\sigma t)-q'\sin(\sigma
t).\ee We see that the classical trajectories have a period of
motion that is energy-dependent and given by $2\pi/\sigma$. If we
remember that $n_0+1/2=(q^2+p^2)/2$ we see that in the
semiclassical limit this time scale becomes precisely $T_c$.

The tangent matrix that is associated with the classical
trajectory that starts in $(q',p')$ can be obtained by simply
differentiating the equations of motion. We must remember that the
angular frequency $\sigma$ is not uniform. The result is
\be\label{tang}
\begin{pmatrix}
  m_{qq} & m_{qp} \\
  m_{pq} & m_{pp}
\end{pmatrix}=\begin{pmatrix}
 \cos(\sigma t) & \sin(\sigma t) \\
 -\sin(\sigma t) & \cos(\sigma t)
\end{pmatrix}\begin{pmatrix}
  1+2q'p't & 2p'^2t \\
  -2q'^2t & 1-2q'p't
\end{pmatrix}.
\ee The action of such a trajectory is easily seen to be
\be\label{skerr}
S=\frac{(\sigma-1)^2T}{4}-i\frac{(\sigma-1)}{2},\ee while the
extra term is \be\label{ikerr} \mathcal{I}=(\sigma-1/2)T.\ee The
result of this semiclassical approximation based on complex
trajectories will be given in section VI.

\subsection{The `leaving' and the `arriving' trajectories}
The first possibility we consider is to approximate the return
probability by using only the real trajectory that leaves the
position $q=0$ with momentum $p$. The tangent matrix for that
trajectory is \be
\begin{pmatrix}
  m_{qq} & m_{qp} \\
  m_{pq} & m_{pp}
\end{pmatrix}=\begin{pmatrix}
 \cos(\sigma t) & \sin(\sigma t) \\
 -\sin(\sigma t) & \cos(\sigma t)
\end{pmatrix}\begin{pmatrix}
  1 & 2p^2T \\
  0 & 1
\end{pmatrix},
\ee which gives the values \be M_{uv}=-ip^2Te^{-i\sigma T}, \quad
M_{vv}=(1+ip^2T)e^{i\sigma T}.\ee The angular frequency is
$\sigma=p^2+1$ and the final points in phase space are
$q_f=p\sin(\sigma T)$ and $p_f=p\cos(\sigma T)$. This corresponds
to \be u''_r=(v''_r)^\ast=\frac{ip}{\sqrt{2}}e^{-i\sigma T}.\ee

Inserting all this information, together with (\ref{skerr}) and
(\ref{ikerr}), into the formula (\ref{Kq1p1}) we have
\begin{align}\label{kq1}
K_{q_1p_1}(z_1,T)&=\frac{1}{\sqrt{1+ip^2T}}\exp\left\{\frac{iT}{4}(p^4+2p^2)-ip^2e^{-i\sigma
T/2}\sin(\sigma
T/2)\right\}\nonumber\\&\times\exp\left\{\frac{ip^4T}{(1+ip^2T)}e^{-i\sigma
T}\sin^2(\sigma T/2) \right\}.\end{align} The first thing we note
is that for $p=0$ we obtain the exact result $K=1$. Moreover, in
the short time regime we can expand (\ref{kq1}) and obtain \be
|K_{q_1p_1}(z_1,T)|^2\approx
1-\frac{1}{2}\left(p^6+3p^4+p^2\right)T^2 \quad \text{(short
times)},\ee which again reproduces the exact calculation. For
later times we should not expect exact agreement.

Let us now turn to the `arriving' trajectory, the one that starts
in $q_i,p_i$ and arrives at te position $q=0$ with momentum $p$
after a time $T$. The tangent matrix is less trivial than in the
previous case, but in the end we get \be M_{vu}=-ip^2Te^{-i\sigma
T}, \quad M_{vv}=(1+ip^2T)e^{i\sigma T}.\ee Since $q^2+p^2$ is a
conserved quantity, we have $\sigma=q_i^2+p_i^2+1=p^2+1$. After
the whole calculation is done, we find out that
$K_{q_2p_2}=K_{q_1p_1}$. This indicates that perhaps these two
approximations will always have the same content of information,
something that is not completely unexpected because of the dual
role of $|z_1\rangle$ and $|z_2\rangle$ in the quantum propagator.

\subsection{The $q_1\to q_2$ possibility}

In that case the trajectories that enter the approximation have
initial momentum given by the equation \be\label{pi}
p_i\sin[(p_i^2+1)T]=0.\ee Of course one solution to this equation
is \be p_i(=p_f)=0,\ee in which case the particle simply stay
still and $\sigma=1$. It is easy to see that for this trajectory
the tangent matrix is very simple, \be\begin{pmatrix}
  m_{qq} & m_{qp} \\
  m_{pq} & m_{pp}
\end{pmatrix}=\begin{pmatrix}
 \cos(T) & \sin(T) \\
 -\sin(T) & \cos(T)
\end{pmatrix}, \ee
and therefore $M_1=M_2=M_{vv}=e^{iT}$. The contribution of this
trajectory to the propagator is \be
K_0=\exp\{-ip^2\sin(T/2)e^{-iT/2}\},\ee where we have used
$S_r+\mathcal{I}_r=T/2$. Notice that for $p=0$ we again have the
exact result $K_0=1$. We also note that the function $|K_0|^2$ has
a period of $2\pi$, which of course corresponds to the quantum
revival time.

We now turn to the other solutions of equation (\ref{pi}). They
are of the form \be p_i^2(n)=\frac{2n\pi}{T}-1,\ee which leads to
$\sigma_n=2n\pi/T$. In this case we have less trivial
trajectories, for which the tangent matrix is given by
\be\begin{pmatrix}
  m_{qq} & m_{qp} \\
  m_{pq} & m_{pp}
\end{pmatrix}=\begin{pmatrix}
 1 & 2p_i^2T \\
 0 & 1
\end{pmatrix}, \ee and we see that the prefactor is
$M_{vv}=(1+ip_i^2T)$, while $M_1=M_2=-4p_i^4T^2$. The action and
the extra term are given by $S_r=(p_i^4T-2ip_i^2)/4$ and
$\mathcal{I}_r=(p_i^2+1/2)T$. The coefficients in (\ref{kqq}) are
\be A_1=A_2=1+\frac{ip_i^2T}{2(1+ip_i^2T)}, \quad
A_{12}=-\frac{1}{1+ip_i^2T}.\ee After many simplifications, we
obtain \be\label{soma} K_{q_1q_2}=\sum_{n=0}^\infty K_n,\ee where
the contribution of the trajectory with label $n$ (different from
zero) is given by \be\label{kn}
K_n=\frac{1}{\sqrt{1+ip_i^2T}}\exp\{-\frac{ip_i^2T}{1+ip_i^2T}(p_i-p)^2+\frac{iT}{4}(p_i^4+4p_i^2+2)\}.
\ee

Note that for short times $p_i(n)$ is very large, so $K_n$ becomes
negligible and $K_0$ gives the only contribution. However, it
predicts the initial decay $|K_0|^2\approx 1-2p^2T^2$, which is
very slow compared to the exact calculation (\ref{short}). The two
results agree only for very small values of the momentum $p$.

Concerning the contributions $K_n$, we see that for a given
instant of time the value of $n$ that will contribute the most is
that for which $p_i(n)$ is as close as possible to $p$, because of
the Gaussian decay in (\ref{kn}). If we impose $p_i^2(n)\approx
p^2$ we have $T\approx 2n\pi/(p^2+1)$, which means that the return
probability has a maximum at the classical period, in agreement
with the exact result.

\subsection{The $q_1\to p_2$ possibility}
If we impose that the classical trajectory starts in $q=0$ with
momentum $p_i$ and ends at $q_f$ with momentum $p$, we have
\be\label{trans1} q_f=p_i\sin(\sigma T), \quad p=p_i\cos(\sigma
T), \quad \sigma=p_i^2+1.\ee These transcendental equations have
no explicit solution. If we confine ourselves to the short time
regime, then we can write $p_i\approx p$ and $q_f\approx
p(p^2+1)T$. The tangent matrix is given by \be
\begin{pmatrix}
  m_{qq} & m_{qp} \\
  m_{pq} & m_{pp}
\end{pmatrix}=\begin{pmatrix}
 \cos(\sigma t) & \sin(\sigma t) \\
 -\sin(\sigma t) & \cos(\sigma t)
\end{pmatrix}\begin{pmatrix}
  1 & 2p_i^2T \\
  0 & 1
\end{pmatrix},
\ee and we obtain $M_2=e^{i\sigma T}(1+2ip_i^2T)$ and
$M_3=e^{-i\sigma T}-2p_i^2T$. Substituting this in (\ref{kq1p2}),
we obtain \be |K_{q_1p_2}|^2\approx
1-\left(p^6+\frac{5}{2}p^4+p^2\right)T^2 \quad \text{(short
times)}, \ee which decays faster than the exact result but is a
better approximation than the one obtained using the $q_1\to q_2$
trajectory. We see that the different approximations may lead to
very different results.

\subsection{The $p_1\to q_2$ possibility}
The equations of motion in this case are \be\label{trans2}
0=q_i\cos(\sigma T)+p\sin(\sigma T), \quad p_f=-q_i\sin(\sigma
T)+p\cos(\sigma T), \quad \sigma=q_i^2+p^2+1,\ee while the tangent
matrix is \be
\begin{pmatrix}
  m_{qq} & m_{qp} \\
  m_{pq} & m_{pp}
\end{pmatrix}=\begin{pmatrix}
 \cos(\sigma t) & \sin(\sigma t) \\
 -\sin(\sigma t) & \cos(\sigma t)
\end{pmatrix}\begin{pmatrix}
  1+2q_ipT & 2p^2T \\
  -2q_i^2T & 1-2q_ipT
\end{pmatrix}.
\ee

The situation here regarding solubility of the equations is even
worse than in the previous case. Once again we restrict the
analysis to the short time regime. Then it is possible to write
the first equation as $q_i\approx -p\sigma T/2$ and find \be
q_i\approx
-\frac{1}{2pT}\left(1-\sqrt{1-4p^2T^2(p^2+1)}\right),\ee which we
substitute in the first equation to find $p_f$. Carrying out the
whole calculation will give in the end \be |K_{p_1q_2}|^2\approx
1-\frac{1}{2}\left(p^6+3p^4+p^2\right)T^2 \quad \text{(short
times)},\ee which agrees with the exact result.

\subsection{The $p_1\to p_2$ possibility}
Finally, in the last possibility we have \be\label{boundpp}
q_f=q_i\cos(\sigma T)+p\sin(\sigma T), \quad p=-q_i\sin(\sigma
T)+p\cos(\sigma T), \quad \sigma=q_i^2+p^2+1.\ee In the short time
limit we have again \be q_i\approx
-\frac{1}{2pT}\left(1-\sqrt{1-4p^2T^2(p^2+1)}\right),\ee and the
final result is \be |K_{p_1p_2}|^2\approx 1-\frac{p^4}{2}T^2 \quad
\text{(short times)}.\ee This is kind of intermediate between the
result we found in subsection B and that of subsections C and D.

\section{Application to a nonlinear oscillator: numerical results}
Before we consider the semiclassical approximations based on real
trajectories for longer times, let us see how well the original
one (\ref{propag2}) compares to the exact result. This has been
considered in detail in [\onlinecite{kerr}], so we just present
the result. Given the initial condition $u'=z_1$, for each time
$T$ we must find a value for $v'$ such that
$v''=z_2^\ast=z_1^\ast$. This problem usually has more than one
solution, and we must add their contributions coherently. In Fig.1
we see the return probability as a function of time (in units of
$T_c$) for the case $p=10$, which we have chosen to ensure that we
are in the semiclassical limit. The corresponding classical period
is $T_c\approx 0.062$. The exact and the semiclassical results are
indistinguishable in this scale.

In the previous section we saw how the different approximations
based on real trajectories performed in the short time regime. The
exact result was reproduced only by the `leaving' and the
`arriving' formulas and by $K_{p_1q_2}$. We now turn to the less
simple case of arbitrary $T$, when the classical trajectories and
the associated propagators must be computed numerically.

Let us start with the propagator $K_{q_1p_1}$, which is based on a
real periodic orbit. Its initial decay is exact, and we can see
how well it does for later times in Fig.2. It is able to reproduce
the height of the peaks with great accuracy, but not their widths.
Since there are never more than one contribution for each time, it
never displays any interference effects.

This is not the case for $K_{q_1q_2}$. We see from (\ref{soma})
and (\ref{kn}) that it consists in the sum of many contributions.
We focus on the values $n=1,2,3$. Their individual contributions
are depicted in Fig.3. Notice that the second and third peaks
overlap. When we calculate the total propagator, this gives rise
to interference. The final result is indistinguishable from the
exact one (for $T>T_c/2$, because we have ignored $K_0$ which
gives a bad initial decay).

So far the propagators could be obtained analytically. Since the
calculation of $K_{q_1p_2}$ depends on the solution of the
transcendental equation (\ref{trans1}), we must resort to
numerical routines. Let us try to find solutions to the second
equation in (\ref{trans1}) in the vicinity of the first period,
$T\approx T_c$. In Fig.4a we see that there are two solutions
(solid lines) for $T<T_c$ and no solution at all for $T>T_c$. This
is because the cosine function with a real argument is always less
than unity, and thus $p_i$ must always be greater than $p$. The
complex solutions do not have this obstruction, as we can also see
in Fig.4a (dashed line), where we plot the real part of the
complex momentum that satisfies the boundary conditions
(\ref{boundary}). Therefore the semiclassical approximation based
on the complex trajectory can reproduce the whole peak, while
$K_{q_1p_2}$ is discontinuous.

In Fig.4b we see the squared modulus of the exact propagator and
the values of $|K_{q_1p_2}|^2$ obtained using the two available
real trajectories. Note that one should not add these results.
They are independent and we may choose any of them, because both
real trajectories are good approximations to the actual complex
one (the real trajectories do not come from a saddle point
approximation). As observed in [\onlinecite{aguiar,novaes}], the
mixed propagator $\langle {\bf x}|e^{-iHT/\hbar}|{\bf z}\rangle$
can also be discontinuous when calculated using real trajectories.
But in that case there are caustics involved, while here we have
an algebraic obstruction.

The discussion of the approximation $K_{p_1q_2}$ is quite similar
to the above. The solutions to the first equation in
(\ref{trans2}) are shown in Fig.5a, where we also show the real
part of the complex position that satisfies the boundary
conditions (\ref{boundary}). Again there is no solution for
$T>T_c$ and the semiclassical propagator is discontinuous, as we
appreciate in Fig.5b. The results are practically the same as in
Fig. 4b.

Finally, the propagator $K_{p_1p_2}$. This time we solve
numerically the conditions (\ref{boundpp}) and find that there is
a single real trajectory for $T<T_c$ and no one for $T>T_c$. The
final result is in Fig.6.

\section{Conclusions}
Several approximations to the semiclassical coherent state
propagator $\langle z_2|e^{-iHT/\hbar}|z_1\rangle$ were presented
that are based solely on real classical trajectories. Two of these
approximations do not involve mixed boundary conditions and thus
are not hindered by the associated `root search' problem. The
remaining four possibilities are based on trajectories that are
determined by initial and final data, but since they are real for
all times they are simpler to determine than the original complex
ones.

As a testing ground we have used the nonlinear system $H=(a^\dag
a)^2$. Only one of the approximations, namely $K_{q_1q_2}$,
reproduced the exact result to the fine details. This is certainly
due to the particular initial coherent state that was chosen, one
corresponding to $q=0$ and $p=10$. Had we chosen for example
$q=10$ and $p=0$ and then $K_{p_1p_2}$ would give excellent
results. We could also consider a nondiagonal propagator, and in
that case we would expect $K_{q_1p_2}$, for example, to improve
its performance.

Straightforward extensions of this work include the already
mentioned case of different position uncertainties (squeezed
states) and also higher dimensional systems. It is also possible
to fix the time $T$ and the initial state $|z_1\rangle$ and to
regard $|K(z_1,z_2,T)|^2$ as a Husimi function defined in the
$z_2$ plane. This is technically more difficult than what we have
presented here, because it involves finding classical trajectories
--usually more than one-- parametrized by points in the plane.

Similar results can be obtained for the semiclassical $SU(2)$, or
spin, coherent state propagator, even though the introduction of
position and momentum variables in that case is not as natural.
The associated phase space is also two-dimensional, but since it
has curvature the calculations may be a little more involved. The
same may be said about the semiclassical $SU(1,1)$ coherent state
propagator. Since these groups have wide applications, it would be
interesting to also have the corresponding approximations based on
real trajectories.

\begin{acknowledgments}
Financial support from FAPESP (Funda\c{c}\~ao de Amparo \`a
Pesquisa do Estado de S\~ao Paulo) is gratefully acknowledged. I
also thank M.A.M. de Aguiar, A.D. Ribeiro and F. Parisio for
important discussions.
\end{acknowledgments}

\pagebreak

\begin{figure}[t]
\includegraphics[scale=0.3,angle=-90]{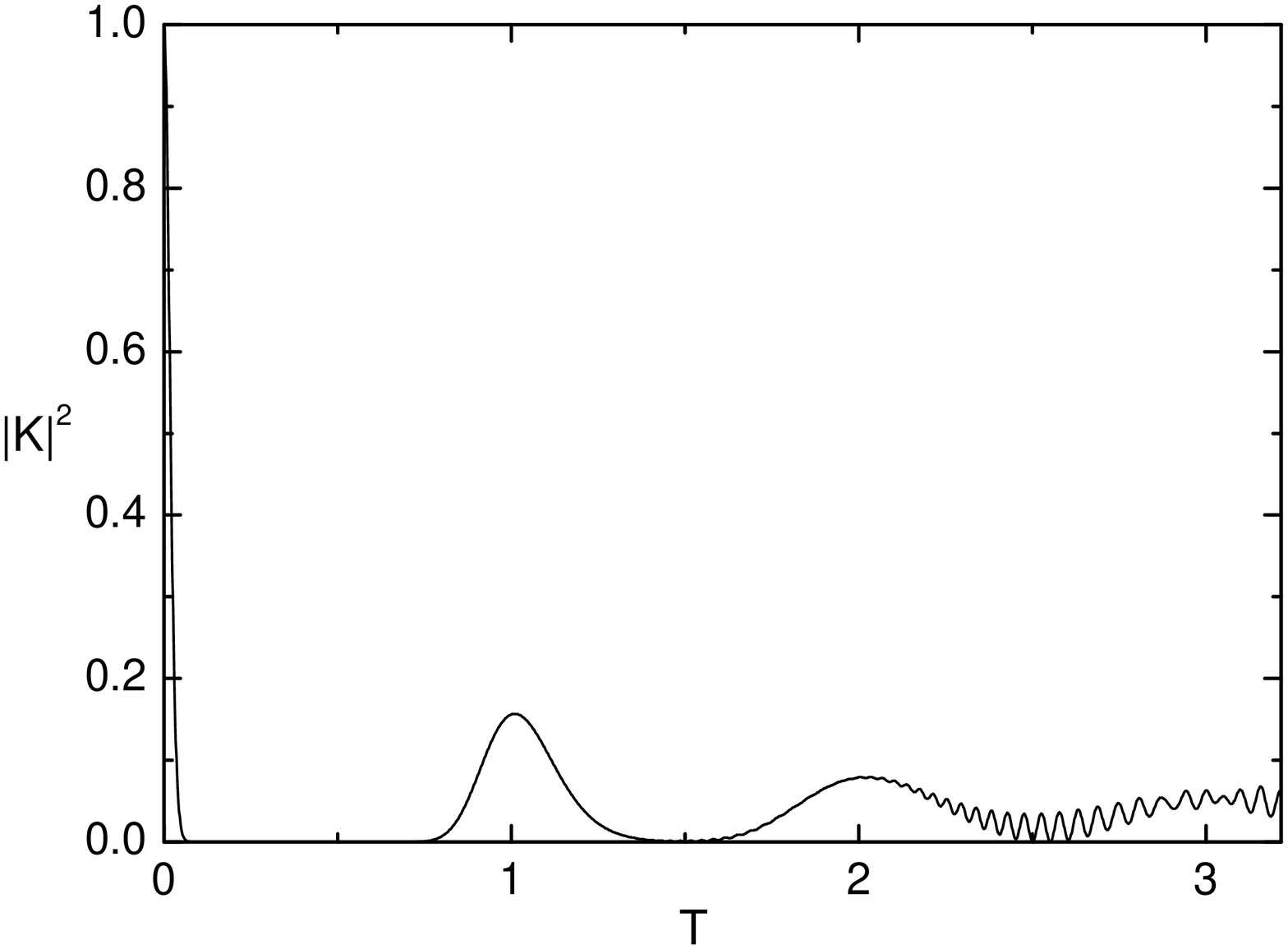}
\caption{Squared modulus of the exact propagator for $q=0$ and
$p=10$. The semiclassical approximation based on complex
trajectories is indistinguishable from it in this scale. Time is
in units of the classical period. }
\end{figure}

\begin{figure}
\includegraphics[scale=0.3,angle=-90]{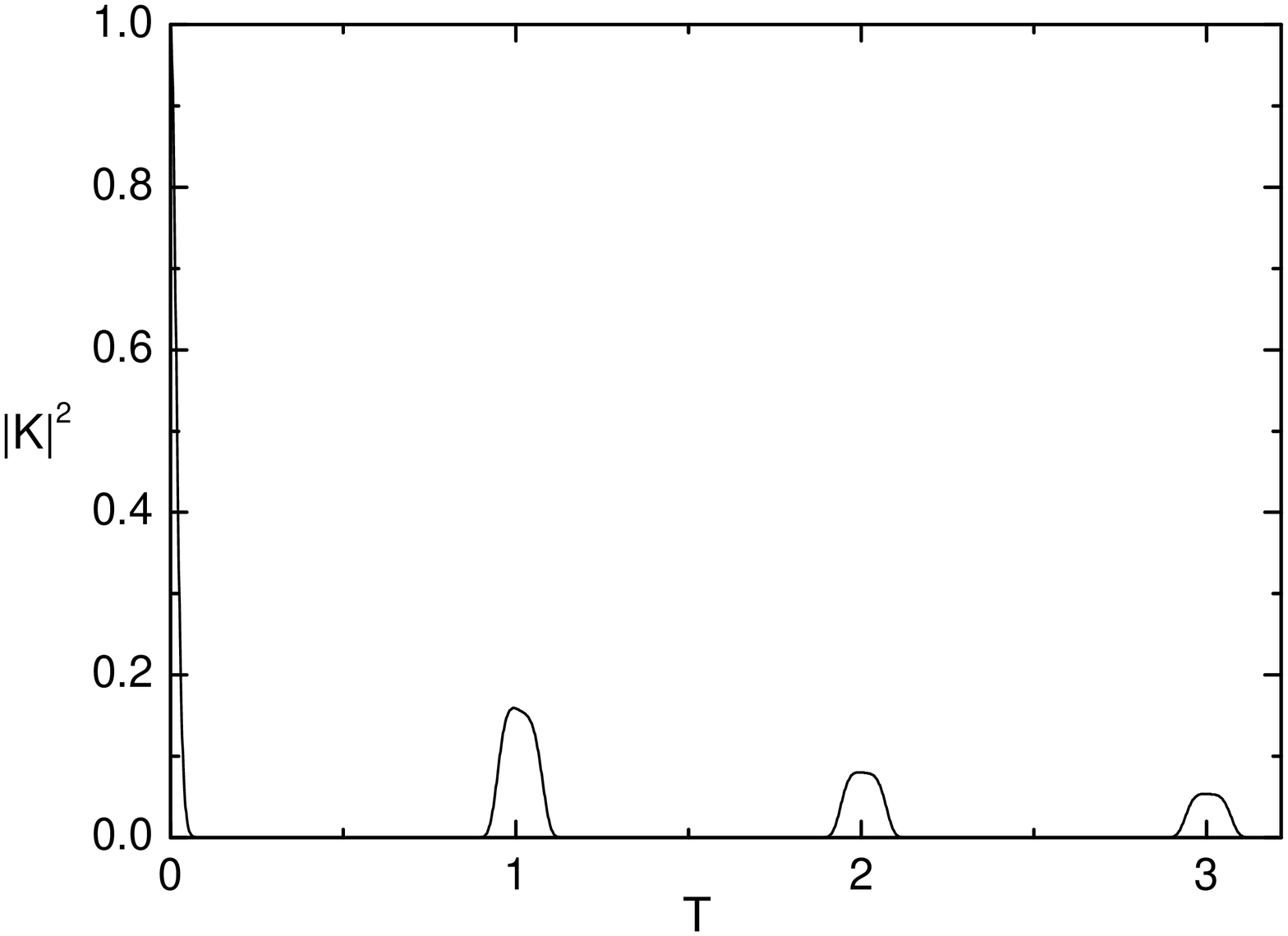}
\caption{The function $|K_{q_1p_1}|^2$ as a function of time. It
reproduces well the height of the peaks, but not their widths, and
shows no interference.}
\end{figure}

\begin{figure}
\includegraphics[scale=0.3,angle=-90]{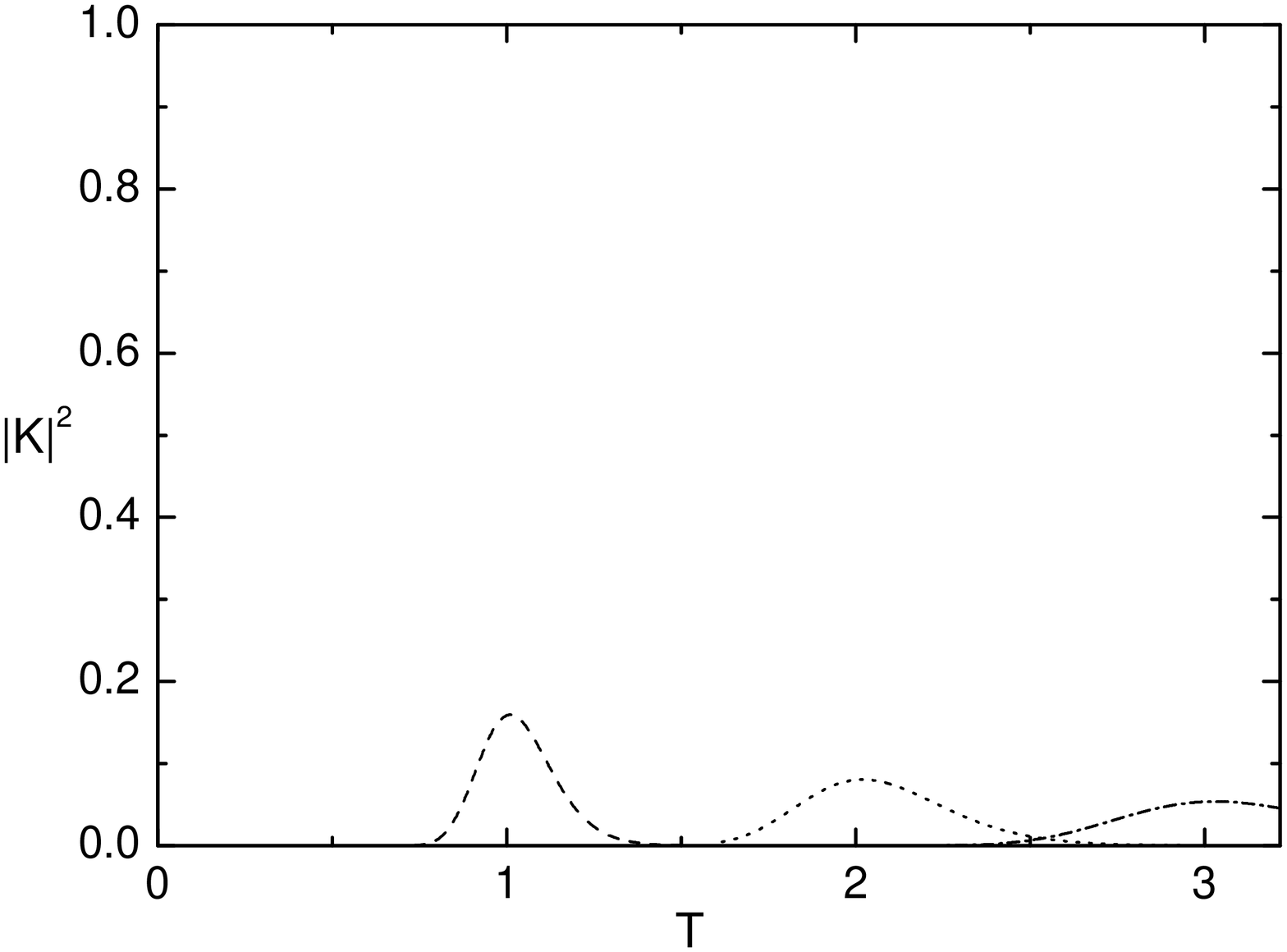}
\caption{The individual contributions $|K_n|^2$ for the
approximation $K_{q_1q_2}$. We show the cases $n=1$, $2$ and $3$.
When they are added there is interference, and the exact result of
Fig.1 is reproduced with extraordinary accuracy for $T>T_c/2$ (we
have not included $K_0$ in the calculation).}
\end{figure}

\begin{figure}
\includegraphics[scale=0.3,angle=-90]{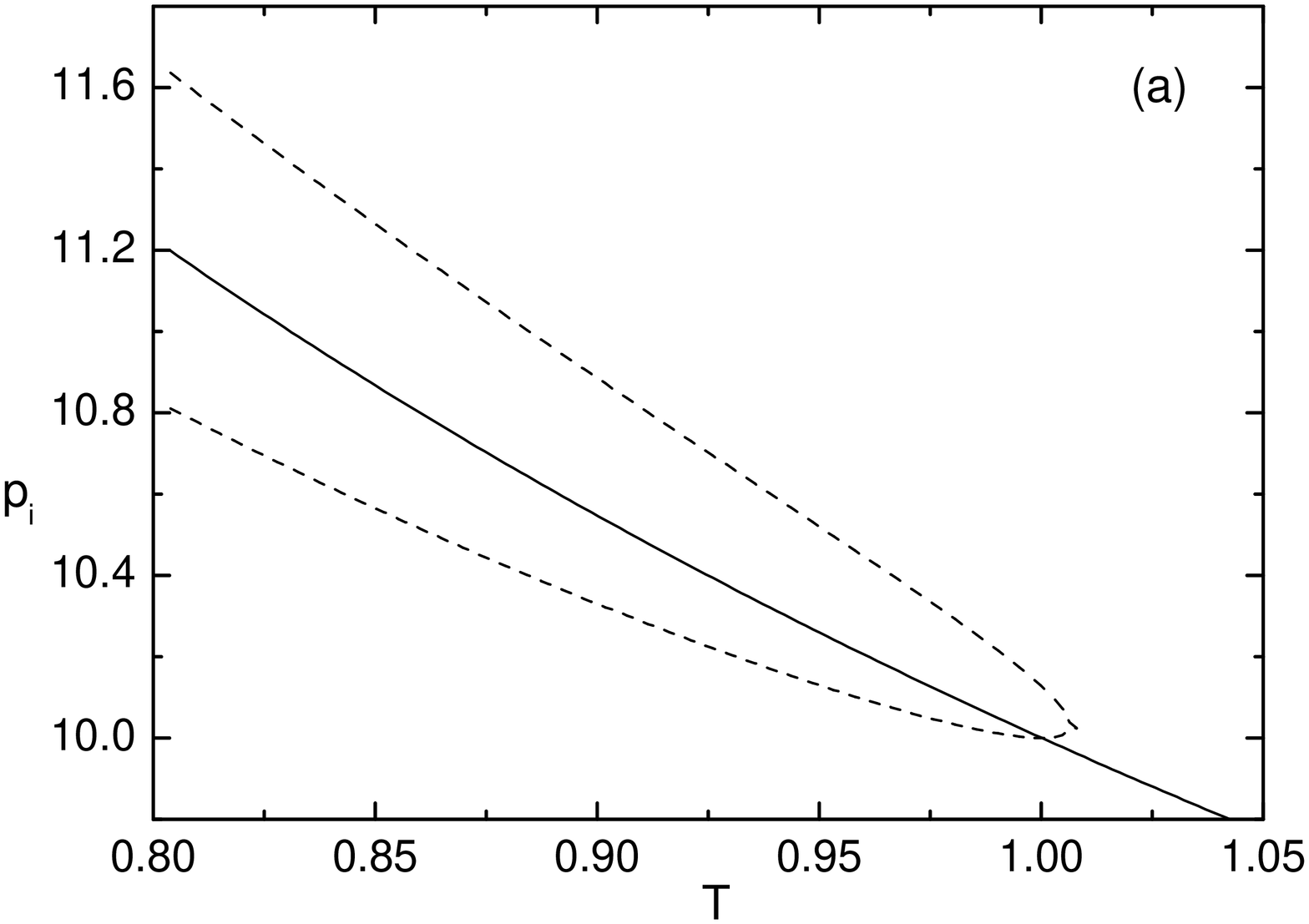}
\includegraphics[scale=0.3,angle=-90]{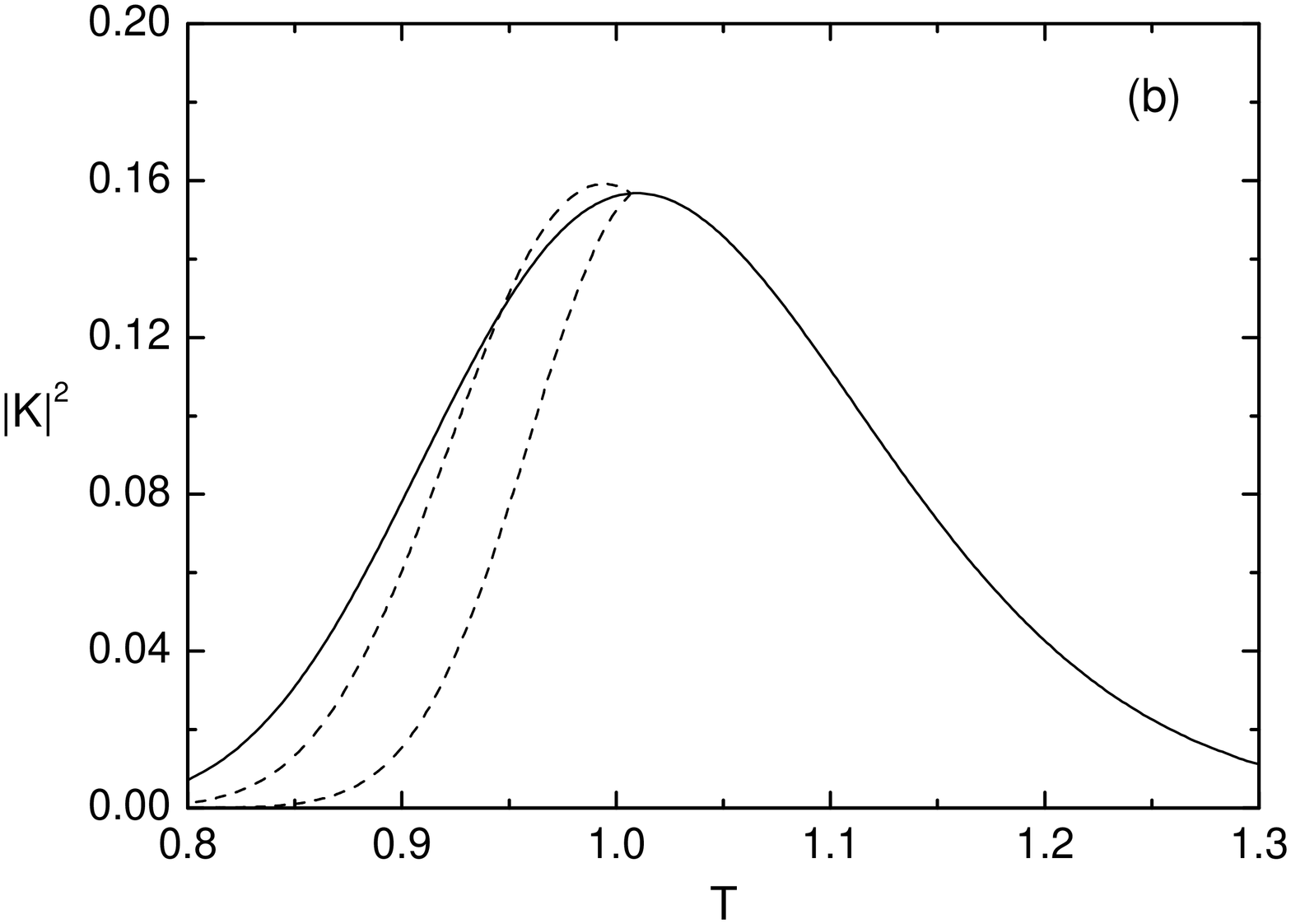}
\caption{Top: real solutions to the equation
$p=p_i\cos((p_i^2+1)T)$ in the vicinity of the classical period
(dashed lines). We also show the real part of the momentum for the
complex trajectory (solid line). Bottom: approximation
$|K_{q_1p_2}|^2$ (dashed lines) compared to the exact result
(solid line). Since there are no real trajectories for $T>T_c$,
the propagator becomes truncated. For $T<T_c$ there are two
possibilities.}
\end{figure}

\begin{figure}
\includegraphics[scale=0.3,angle=-90]{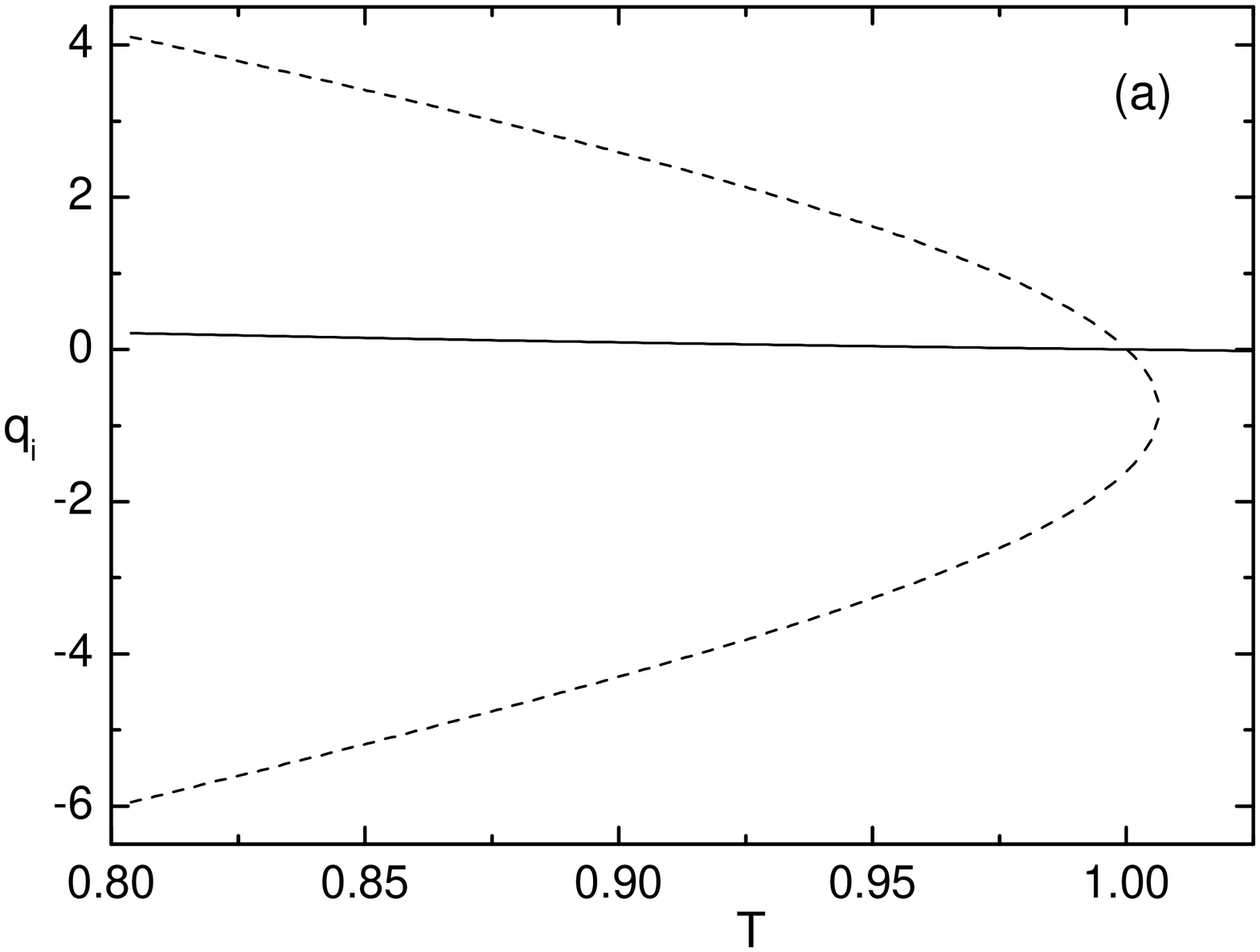}
\includegraphics[scale=0.3,angle=-90]{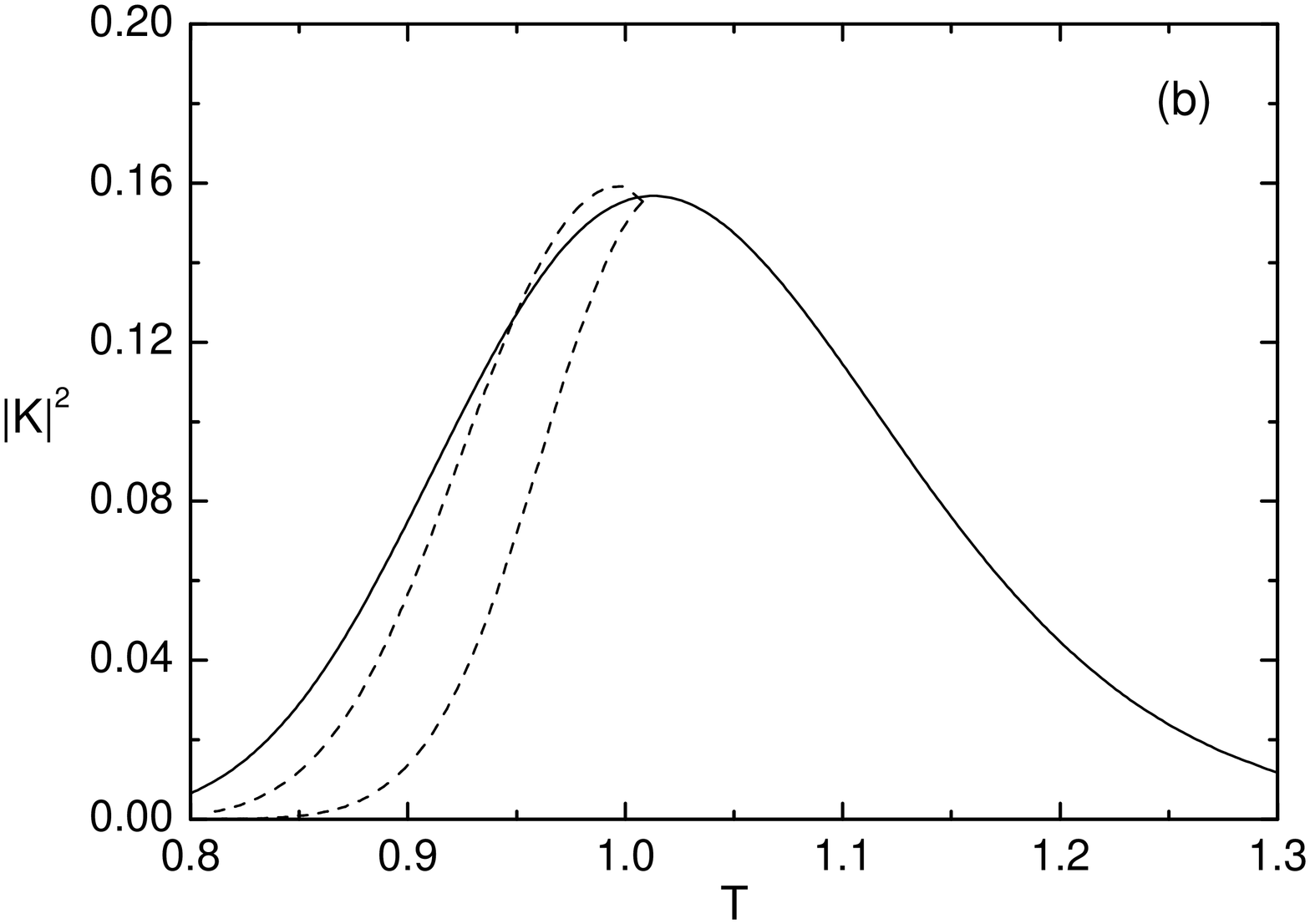}
\caption{Top: real solutions to the equation $0=q_i\cos(\sigma
T)+p\cos(\sigma T)$, where $\sigma=q_i^2+p^2+1$, in the vicinity
of the classical period (dashed lines). We also show the real part
of the position for the complex trajectory (solid line). Bottom:
approximation $|K_{p_1q_2}|^2$ (dashed lines) compared to the
exact result (solid line). The situation is analogous to the
previous figure.}
\end{figure}

\begin{figure}[t]
\includegraphics[scale=0.3,angle=-90]{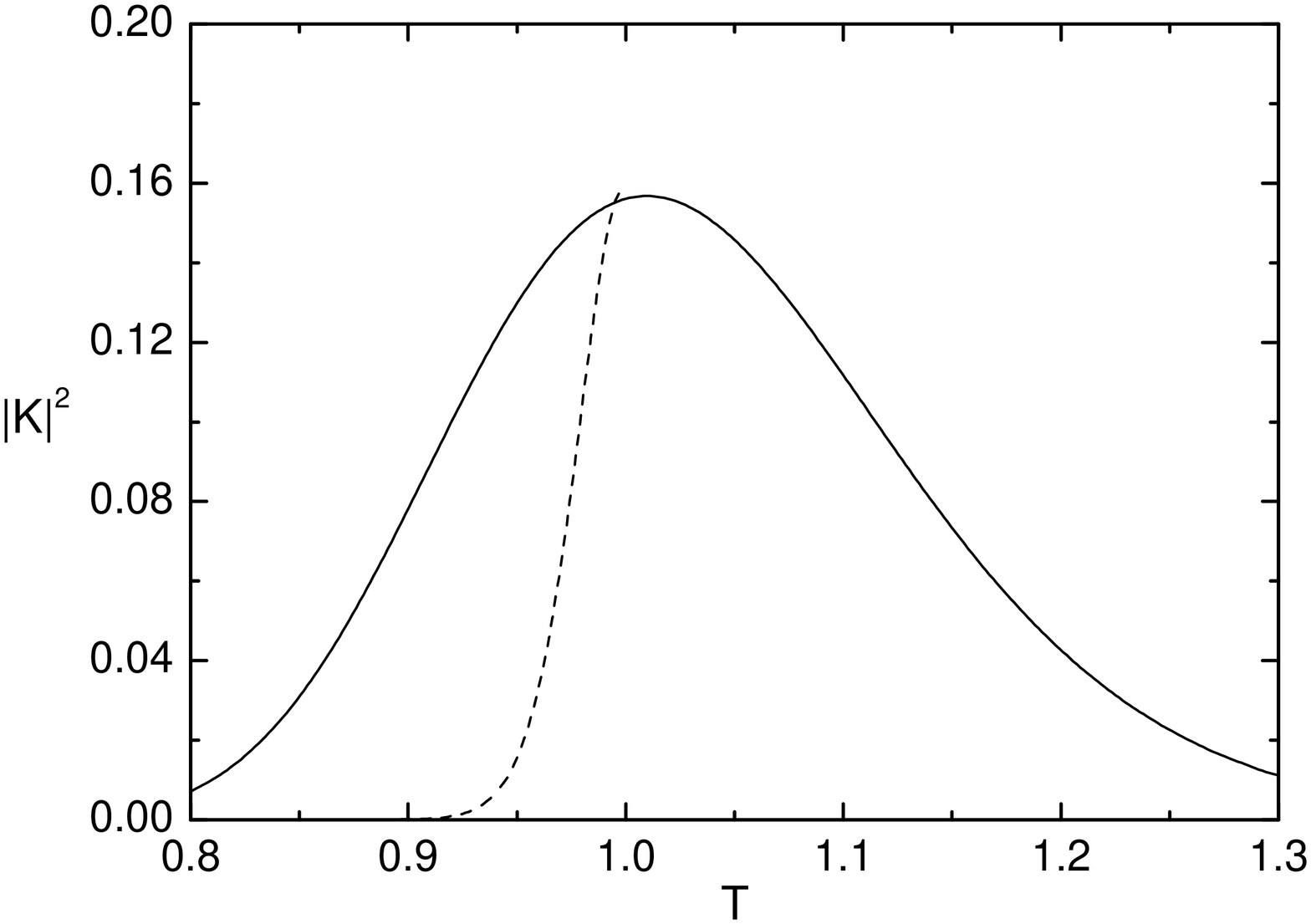}
\caption{Approximation $|K_{p_1p_2}|^2$ (dashed line) compared to
the exact result (solid line). This time only one real trajectory
exists for $T<T_c$, but again the propagator is truncated.}
\end{figure}


\begin{thebibliography}{10}

\bibitem{klauder} J.R. Klauder, in {\it Path Integrals} (G.J. Papadopoulos and J.T.
Devreese, Eds.), Plenum, New York, 1978, p.5; Phys. Rev. D {\bf
19}, 2349 (1979); Phys. Rev. Lett. {\bf 56}, 897 (1986); in {\it
Random Media} (G. Papanicolau, Ed.), Springer-Verlag, New York,
1987, p.163.

\bibitem{rubin} A. Rubin and J.R. Klauder, Ann. Phys. (N.Y.) {\bf 241},
212 (1995).

\bibitem{weiss} Y. Weissman, J. Chem. Phys. {\bf 76}, 4067 (1982); J. Phys. A {\bf 16}, 2693
(1983).

\bibitem{unpub} M.A.M. de Aguiar and M. Baranger, 1989, unpublished.

\bibitem{baranger} M. Baranger {\it et al}, J. Phys. A {\bf 34}, 7227 (2001).
\bibitem{solari} H.G. Solari, J. Math. Phys. \textbf{28}, 1097
(1987); E.A. Kochetov, {\it ibid} \textbf{36}, 4667 (1995); V.R.
Vieira and P.D. Sacramento, Nucl. Phys. B \textbf{448}, 331
(1995); E.A. Kochetov, J. Phys. A {\bf 31}, 4473 (1998); M. Stone,
K.-S. Park and A. Garg, J. Math. Phys. \textbf{41}, 8025 (2000);
K.-S. Park, M. Stone and A. Garg, Int. J. Mod. Phys. B {\bf 15},
3220 (2001); M. Pletyukhov, J. Math. Phys. {\bf 45}, 1859 (2004).

\bibitem{adachi} S. Adachi, Ann. Phys. (N.Y.) {\bf 195}, 45
(1989).

\bibitem{marcus} A.L. Xavier Jr and M.A.M. de Aguiar, Phys. Rev. A
{\bf 54}, 1808 (1996); Ann. Phys. (N.Y.) {\bf 252}, 458 (1996).

\bibitem{xavier} A.L. Xavier Jr and M.A.M. de Aguiar, Phys. Rev. Lett. {\bf 79}, 3323 (1997).

\bibitem{gross} F. Grossmann, Phys. Rev. A {\bf 57} 3256 (1998).

\bibitem{voorhis1} T. Van Voorhis and E.J. Heller, Phys. Rev. A
{\bf 66}, 050501 (2002).

\bibitem{voorhis2} T. Van Voorhis and E.J. Heller, J. Chem. Phys. {\bf 119}, 12153 (2003).

\bibitem{cabelo} A.D. Ribeiro, M.A.M. de Aguiar and M. Baranger,
Phys. Rev. E {\bf 69}, 66204 (2004).

\bibitem{spin} E. Kececioglu and A. Garg, Phys. Rev. Lett. {\bf 88}, 237205 (2002);
Phys. Rev. B {\bf 67}, 054406 (2003); M. Novaes, {\rm
quant-ph/0505224}.

\bibitem{huber} D. Huber and E.J. Heller, J. Chem. Phys. {\bf 87}, 5302 (1987);
D. Huber, E.J. Heller and R. Littlejohn, {\it ibid} {\bf 89}, 2003
(1988).

\bibitem{aguiar} M.A.M. de Aguiar {\it et al}, J. Phys. A {\bf 38}, 4645 (2005).
\bibitem{parisio1} F. Parisio and M.A.M. de Aguiar, to appear in J. Phys.
A.

\bibitem{novaes} M. Novaes and M.A.M. de Aguiar, {\rm quant-ph/0504037}, to appear in Phys. Rev.
A.

\bibitem{cellular} E.J. Heller, J. Chem. Phys. {\bf 94}, 2723
(1991).

\bibitem{houches} E.J. Heller in {\it Chaos and Quantum Physics},
edited by M.J. Giannoni, A. Voros and J. Zinn-Justin, Les Houches
Session LII, 1989 (Elsevier, Amsterdam, 1991).

\bibitem{stadium} S. Tomsovic and E.J. Heller, Phys. Rev. Lett. {\bf 67}, 664
(1991); S. Tomsovic and E.J. Heller, Phys. Rev. E {\bf 47}, 282
(1993).

\bibitem{cel2} M.A. Sep\'ulveda, S. Tomsovic and E.J. Heller, {\it ibid} {\bf
69}, 402 (1992);

\bibitem{gutz} J.H. Vleck, Proc. Natl. Acad. Sci. U.S.A. {\bf
14}, 178 (1928); M.C. Gutzwiller, J. Math. Phys. {\bf 8}, 1979
(1967).

\bibitem{HK} M.F. Herman and E. Kluk, Chem. Phys. {\bf 91}, 27
(1984).

\bibitem{review} W.H. Miller, J. Phys. Chem. A {\bf 105}, 2942
(2001); M. Thoss and H. Wang, Ann. Rev. Phys. Chem. {\bf 55}, 299
(2004); K.G. Kay, {\it ibid} {\bf 56}, 255 (2005).

\bibitem{uniform} A.D. Ribeiro, M. Novaes and M.A.M. de Aguiar, {\rm quant-ph/0505155}, to appear in Phys. Rev.
Lett.

\bibitem{parisio} F. Parisio and M.A.M. de Aguiar, Phys. Rev. A {\bf 68}, 062112
(2003).

\bibitem{revivals} R. Bluhm, V.A. Kostelecky and J.A. Porter, Am. J. Phys. {\bf 64},
944 (1996).
\bibitem{kerr} M. Novaes and F. Parisio, to appear.

\end{thebibliography}
\end{document}